\documentclass[sigconf]{acmart}

\def\BibTeX{{\rm B\kern-.05em{\sc i\kern-.025em b}\kern-.08emT\kern-.1667em\lower.7ex\hbox{E}\kern-.125emX}}
    
\copyrightyear{2019}
\acmYear{2019}
\acmConference[MMAsia '19]{ACM Multimedia Asia}{December 15--18, 2019}{Beijing, China}
\acmBooktitle{ACM Multimedia Asia (MMAsia '19), December 15--18, 2019, Beijing, China}

\usepackage{float}
\usepackage{tabularx}
\usepackage{array}
\newcolumntype{C}[1]{>{\centering\let\newline\\\arraybackslash\hspace{0pt}}m{#1}}
\usepackage{graphicx}
\usepackage{CJKutf8}
\usepackage{multirow}
\usepackage{url}
\usepackage{hyperref}
\begin{document}

%
\title{Measuring Similarity between Brands \\ using Followers' Post in Social Media}
\renewcommand{\shorttitle}{Measuring Similarity between Brands using Followers' Post in Social Media}

%

\author{Yiwei Zhang}
\email{zhangyiwei@hal.t.u-tokyo.ac.jp}
\affiliation{%
  \institution{The University of Tokyo}
}

\author{Xueting Wang}
\email{xt\_wang@hal.t.u-tokyo.ac.jp}
\affiliation{%
	\institution{The University of Tokyo}
}
\author{Yoshiaki Sakai}
\email{sakai.yoshiaki@geomarketing.co.jp}
\affiliation{%
	\institution{Geomarketing Co.,Ltd.}
}

\author{Toshihiko Yamasaki}
\email{yamasaki@hal.t.u-tokyo.ac.jp}
\affiliation{%
	\institution{The University of Tokyo}
}

%

%
\begin{abstract}
In this paper, we propose a new measure to estimate the similarity between brands via posts of brands' followers on social network services (SNS). Our method was developed with the intention of exploring the brands that customers are likely to jointly purchase. Nowadays, brands use social media for targeted advertising because influencing users' preferences can greatly affect the trends in sales. We assume that data on SNS allows us to make quantitative comparisons between brands. Our proposed algorithm analyzes the daily photos and hashtags posted by each brand's followers. By clustering them and converting them to histograms, we can calculate the similarity between brands. We evaluated our proposed algorithm with purchase logs, credit card information, and answers to the questionnaires. The experimental results show that the purchase data maintained by a mall or a credit card company can predict the co-purchase very well, but not the customer's willingness to buy products of new brands. On the other hand, our method can predict the users' interest on brands with a correlation value over 0.53, which is pretty high considering that such interest to brands are high subjective and individual dependent.
\end{abstract}


%

%

%
\maketitle

\section{Introduction}
Billions of photographs are uploaded to the Internet every day through various photo \& video sharing services. Social media is already a part of our everyday life.
The rapidly increasing popularity of social media has given rise to new marketing approaches.
Social media marketing is the use of social media platforms and websites to promote a product or service \cite{felix2017elements}.
Companies make use of platforms such as Twitter\footnote{\url{https://twitter.com}} and Instagram\footnote{\url{https://www.instagram.com}} to reach a much wider spectrum of audience than that of traditional advertisements such as newspaper, television, and radio.
By 2018, 91\% of the Fortune 500 companies were actively using Twitter, and 63\% of them had corporate Instagram accounts\footnote{\url{https://www.umassd.edu/cmr/social-media-research/2018-fortune-500}}.
With these newfound trends, we can use data from social media to analyze customers' preferences and the implicit relationship between brands.
Understanding the relationship between brands, products, and people can have a profound effect on marketing. For example, when a brand wants to find celebrities for endorsements, it is important to consider whether his/her followers are potential customers for its products. Similarly, when a shopping mall is in the process of choosing the brands it wishes to carry, considering the specific combinations of brands that have the potential to attract more customers is a lucrative proposition. However, such design or decision making has been done by only experts with sense of trend so far.

Co-purchasing history from credit cards or point cards in a shopping mall can be analyzed to measure the similarity between brands. Unfortunately, there are some demerits of this method.
Some customers, especially teenagers, may not own a credit card. Secondly, people prefer using cash for small purchase amounts. Consequently, brands that offer products in the low price range may seldom appear in the purchasing history.
Many shopping malls have their own point card systems, which allow customers to earn points every time they make a purchase from one of the franchises. But these analyses are limited by the number of brands existing at the location of the purchase. Namely, similarity between brands that are not co-located in the same region can not be calculated.

Because of these reasons, we chose social network data to measure the similarity between brands. Unlike the prerequisites for credit cards or point cards, social media has no age or income requirement, also there are no limitations to the number of brands being advertised by marketers or users.
One might think that SNS penetration is lower among elder generation and therefore, it is hard to get information from them. But according to the investigation\footnote{\url{https://www.pewinternet.org/fact-sheet/social-media}}, 64\% of Americans in the age group 50-64 use SNS actively as well as 37\% in the age group of 65 years and above.

We briefly summarize the main contributions and findings of this study:
\begin{itemize}
	\item We have presented an approach to predict the similarity between brands using tags and images posted by the brands' followers. We show that such off-the-shelf multimedia technologies, though not technically novel, can be a new marketing tool.
	\item We have created an Instagram brand datasets including 233,000 unique users and their most recent 100 posts for brand similarity measurement.
	We release Instagram dataset to the public\footnote{\url{https://github.com/yiwei51/brand_dataset}}.
	We have also created two evaluation datasets based on the purchasing histories of two real-world customers for results comparison.
	\item We have conducted user studies by questionnaires to obtain the users' co-purchasing, interest, and knowledge tendencies. It provides an additional angle of evaluation besides point card and credit card purchasing histories.
\end{itemize}

\section{Related Works}
\subsection{Similarity Measurement}
There are many related studies on measuring latent similarity with multimedia data.
Item2Vec \cite{barkan2016item2vec} measured similarity between item for recommendation. They considered items purchased by one user as a positive example then used a skip-gram model to produce embeddings. Eisner et al. used a similar method to convert an emoji to a vector then measured similarity between emoji \cite{eisner2016emoji2vec}. \cite{liu2014social} measured latent similarity between social images by metric learning. Wang et al. used graph embedding techniques to find similar items \cite{wang2018billion}. Hsiao et al. used the topic models to learn style-coherent representations for fashion images retrieval \cite{hsiao2017learning}.

\subsection{Brand Analysis using Social Media}
Researches on brands are increasingly focused on the content on social media.
Goh et al. showed that engagement in social media brand communities leads to a positive increase in sales \cite{goh2013social}.
In a study of understanding brands' perceptions, Culotta et al. used the data on Twitter to predict how consumers perceive brands' attributes including eco-friendliness, nutrition, and luxury \cite{culotta2016mining}.
Several researchers analyzed the popularity of brands and their social media posts.
De Vries et al. reported that brand posts might be popular due to several cues such as vividness, interactivity, and valence of comments \cite{de2012popularity}. Mazloom et al. investigated similar indicators with text, image and social features on Instagram \cite{mazloom2016multimodal}.

\subsection{Content-based Recommendation}
Recommendation is another topic of interest in the study of social media, both commercially and academically.
Besides widely used collaborative filtering methods \cite{breese1998empirical}, another common approach is content-based recommendation \cite{pazzani2007content}. Apart from users-item interactions, content-based methods analyze information such as user profiles and item relations \cite{xin2019relational}.
Gelli et al. presented a study on image recommendations for brands. They recommended images which were consistent with a brand's style, by learning features from images uploaded by these brands \cite{gelli2018beyond}.

\subsection{Brand Similarity}
The pairwise similarity between brands was measured previously by Bijmolt et al. \cite{bijmolt1998judgments}. They used questionnaires to ask people regarding the likeness of two products to identify the similarities between the products of various brands. The limitation of this method was that it required time and monetary investment to collect sufficient data. In \cite{yangunderstanding}, they aimed to find similar brands in the same category using customers' reviews, while their approach is not suitable for finding the relationship between brands in diverse product categories. 

In \cite{segalin2017your}, the researchers predicted user's personality based on his or her profile picture on Facebook. We also believe that images and hashtags posted by users are indicative of their personalities and users' choice of brands represents their preferences. Different from \cite{bijmolt1998judgments} and \cite{yangunderstanding}, our proposed method could find the relationship between two diverse product categories without tremendous time and monetary investment.

\section{Dataset Construction}
\begin{table*}[!t]
	\begin{center}
		\caption{Summary of datasets used in this paper.} \label{table:brand_dataset}
		\vspace{-0.1cm}
		\fontsize{8.5pt}{10pt}\selectfont
		\begin{tabular}{|c|c|c|c|c|c|c|} \hline
			\multirow{2}{*}{Dataset} & \multirow{2}{*}{\# Brands} & \# Instagram users & \multirow{2}{*}{Data source} & \multirow{2}{*}{\# Transactions} & \multirow{2}{*}{\# Customers} & \multirow{2}{*}{Period of record}\\ 
			&  & (1,000 followers/brand) &  &  & &\\ \hline
			Pointcard100 & 100 & 100,000 & point card used in a mall& 1,759,098 & 241,937 & 2015.04.01 - 2016.03.31\\ \hline
			Creditcard81& 81 & 81,000 & a national credit card brand& 9,393,546 & 1,236,521 & 2016.01.01 - 2017.12.31\\ \hline 
			Popular108 &108 & 108,000 & Instagram&-&-&-\\ \hline
		\end{tabular}
	\end{center}
\end{table*}

\begin{figure}[t!]
	\centering
	\includegraphics[width=0.95\linewidth]{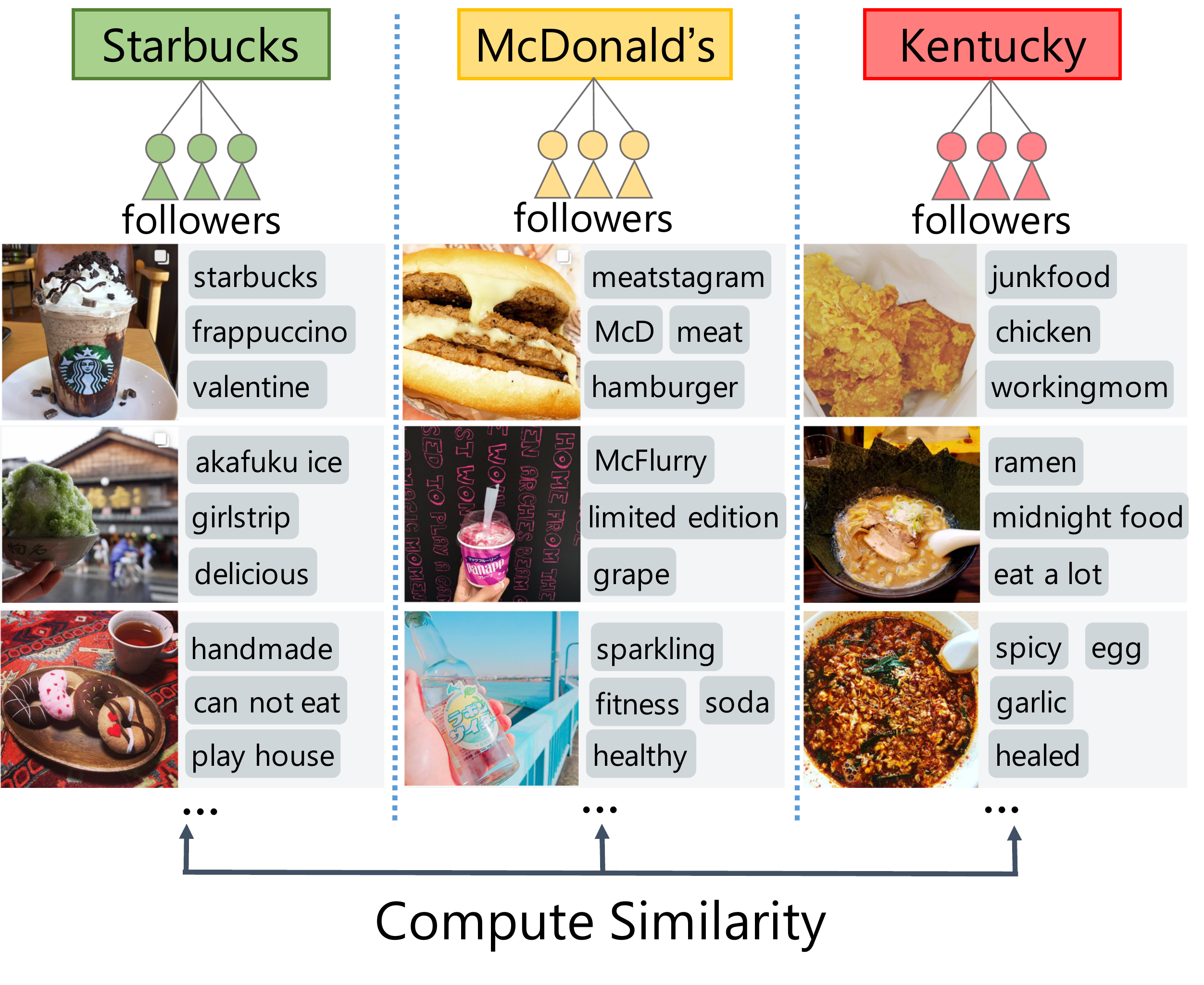}
	\vspace{-0.1cm}
	\caption{Example posts from brand followers.}
	\label{fig:brand_follower}
	\vspace{-0.1cm}
\end{figure}

\begin{figure*}[t!]
	\centering
	\includegraphics[width=0.89\linewidth]{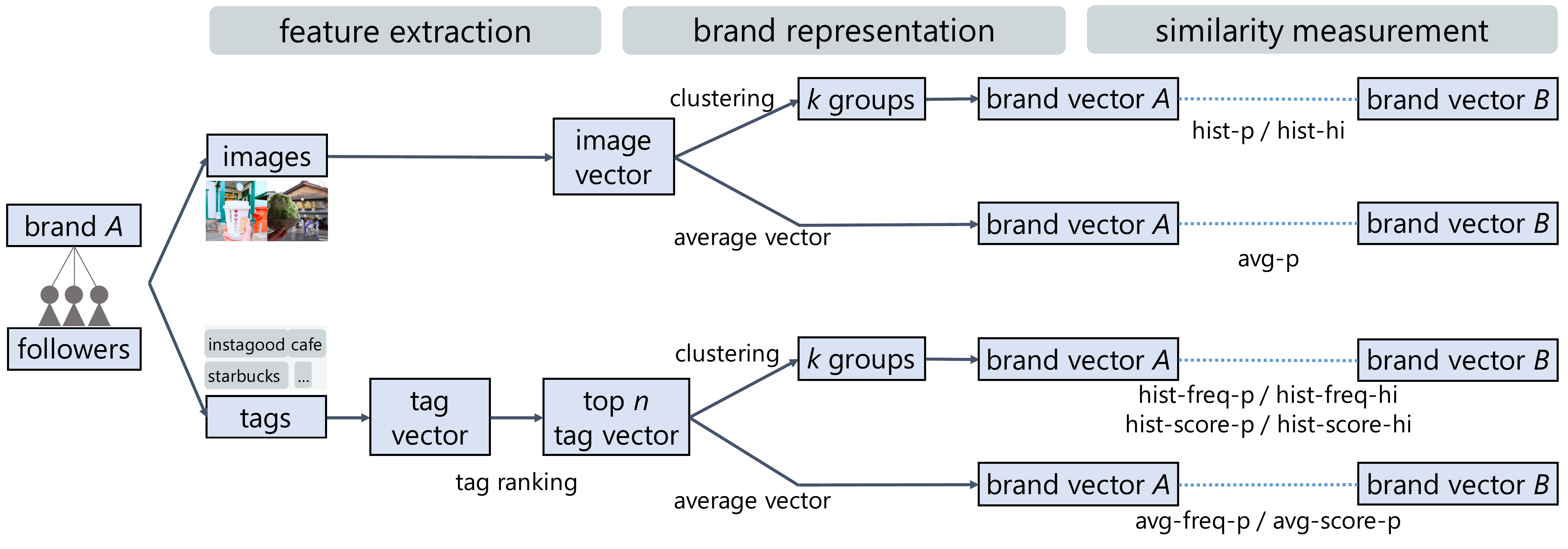}
	\caption{The flowchart for the feature extraction, brand representation and similarity measurement processes.}
	\label{fig:flow}
	\vspace{-0.1cm}
\end{figure*}
Instagram is a social photo sharing service. We chose to collect posts from Instagram because many brands have created official accounts for promoting the brand image.
We show posts from the followers of three brands separately in Figure~\ref{fig:brand_follower}. Though these posts are not from the brands' official accounts, we can see that followers of different brands have different styles. Their posts indicate the everyday activities of these followers, which, we believe, can be utilized to represent the similarity between the brands.
We chose to use followers' posts instead of brands' posts because brands in the same category may have greater similarity based on their posts. For example, a fashion brand will be more similar to another fashion brand than to a cosmetic brand. However, finding the relationship between two diverse product categories is of greater interest to us.
The problem of measuring the pairwise similarity between brands transforms to the problem of measuring the pairwise similarity between their followers. We randomly selected 1,000 followers per brand and used their posts to represent this brand. We chose to associate a unique user to just one brand. We collected data on these followers by web crawling their Instagram posts.

We created three datasets in our study. The difference between the datasets is shown in Table~\ref{table:brand_dataset}.
Pointcard100 and Creditcard81 datasets are created for the evaluation. The Pointcard100 dataset included 100 brands that was obtained from customers' purchasing histories at a shopping mall in Japan. The Creditcard81 dataset included 81 brands and comprised customers' purchasing histories maintained by a credit card company in Japan. They were both anonymized before processing. Both datasets included information on products of specific brands purchased by the users.
The Popular108 dataset included the official accounts of 108 of the most popular companies in Japan. The number of followers in Instagram was directly used to express the popularity.


\section{Proposed Methods}\label{proposedmethods}
We present a new scale for measuring the similarity between brands by analyzing posts from brand followers. A flowchart for the proposed methods consisting of the feature extraction, brand representation and similarity measurement processes is depicted in Figure~\ref{fig:flow}.
\subsection{Feature Extraction}
\subsubsection{Image Feature}
For each brand, we used the 10 most recent photographs posted by 1,000 users to represent the brand.
We used a 50-layer ResNet \cite{he2016deep} pretrained on ImageNet \cite{deng2009imagenet} to extract the image features. In this step, we transform each image into a feature vector of column size 2048.
\subsubsection{Tag Feature}
To extract efficient tag features, we took two steps: (1) tag embedding to covert each tag into a vector representation; (2) tag ranking to select the top ranked tags as the typical tag feature for brand representation.

Tag embedding process is inspired by the work Item2Vec \cite{barkan2016item2vec}. It produces embedding for items in a latent space using Word2Vec. We considered tags belonging to one image as words in one sentence. Then, we used \textit{fasttext}{\texttrademark} to convert each tag to a 100-dimension vector \cite{bojanowski2017enriching}.

In the tag ranking, we select the top 3,000 tags to represent the brand. The first method is to rank by frequency and the second method is to rank by tag score.

\textit{Rank by Frequency:}
In this method, we rank tags by the number of users who have used it.
In \cite{boakye2017tag}, the researchers observed that a large number of users use the same tags to every photo.
Therefore, we decided to rank the tags by the number of users who have used the tag at least once. In other words, regardless of the number of times the user has used the tag, we only counted the tag once for one user.

\textit{Rank by Tag Score:}
Instead of utilizing the number of users who have used the tag at least once, we apply term frequency–inverse document frequency (TF–IDF) \cite{salton1986introduction} for selecting the tags.
TF–IDF is a numerical statistic, which indicates the importance of a word to a document in a collection or corpus \cite{anand2012mining}.
In our tag ranking algorithm, we considered a brand as a document, and tags belong to a brand as words inside a document. We utilized the number of users who have used the tag at least once to represent the TF. The TF score is higher if more people have used the tag.
In contrast, we consider that the IDF score would be lower if the tag has been used for other brands frequently. 
The score for the tag is calculated using the following equation:
\begin{align}
	tag\_score(t_i, b_j) &= tf(i, j) \times idf(i) , \\
	tf(i, j) &= \sum_{k} \frac{n_{i, j}}{n_{k, j}} \label{eq:tag_tf} , \\
	idf(i) &= \log{\frac{B}{\sum_{j} c_{i, j}}}+ 1 \label{eq:tag_idf} , \\
	c_{i, j} &= \begin{cases}
		1& \text{$r_{i, j}$ $\leq$ $l$}\\
		0& \text{$r_{i, j}$ $>$ $l$} ,
	\end{cases}
\end{align}
where $n_{i,j}$ is the number of users that have used the tag $t_i$ in brand $b_j$, and $B$ is the number of brands. 
$r_{i,j}$ is the frequency ranking of tag $t_i$ in the brand $b_j$, $c_{i,j}$ is 1 if the tag $t_i$ appears in the top $l$ tags of brand $b_j$, else $c_{i,j}$ is 0. We set $l = 1,000$ in this experiment.

\subsection{Brand Representation}
After converting the images and tags into vectors, we addressed the problem of representing a brand using these vectors.
In the following section, we introduce several ways to handle image and tag features, including histogram-based methods and average vector based methods, and compare their performances.

\textit{Histogram-based Method:}
We used mini-batch K-means to cluster features after feature extraction to form a bag of features. We have tried using $K$ from 10 to 10,000 and employed the best one.
After clustering, we use the number of images/tags in each cluster as the brand vector.

\textit{Average Vector based Method:}
In the case of images, we average the image vectors directly and use it as the brand vector.
In the case of tags, after tag ranking, we select top 3,000 tags and average their tag vectors, then we use this averaged vector as the brand vector.


\subsection{Similarity Measurement}
\begin{figure*}[t!]
	\centering
	\includegraphics[width=0.87\linewidth]{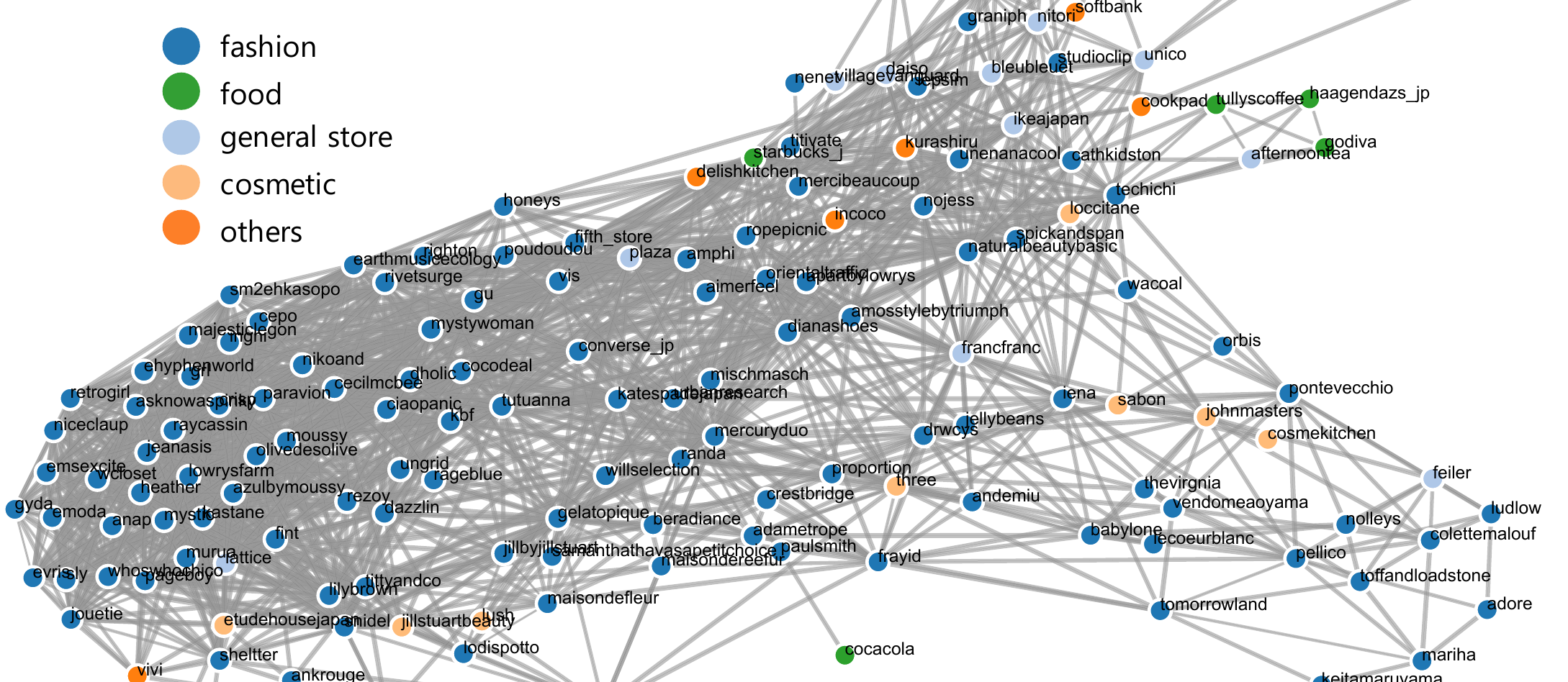}
	\vspace{-0.2cm}
	\caption{Visualization of the brand relationships (full version of the visualization\protect\footnotemark).}
	\label{fig:plot}
\end{figure*}

We considered two ways for histogram-based methods in similarity measurement. The first is the Pearson correlation and the second is histogram intersection similarity. 
For the average vector based methods, we only considered the Pearson correlation.
We calculated similarity between each pair of brand vectors.
Figure~\ref{fig:plot} shows the visualization of the brand relationships based on the histogram method of ranking by the tag score. We can easily find relationship between brands from this figure. For example, for those who like ``Starbucks,'' they might also be interested in fashion brands ``titivate.''

\vspace{-0.2cm}

\section{Evaluation}
\subsection{Stability}
Because our proposed methods are based on 1,000 followers randomly selected for each brand, we conducted an experiment to evaluate the stability of our proposed methods.
For each brand, we randomly split the 1,000 followers into two groups for 5 times and then computed the similarity between the brands for each group. 
The average Spearman's ranking correlation coefficient based on the two groups was calculated as 0.98. It indicates that our proposed methods are stable.

We also conducted an experiment to show that less number of followers might be enough for our purpose. We randomly selected a small group of followers from each brand for 5 times and then computed the similarity between using a small group of followers and using 1,000 followers. We found that the average Spearman's ranking correlation coefficient was calculated as 0.93 when using 100 followers per brand, and 0.88 when using 50 followers per brand. We think 100 followers might be enough for our purpose.


\subsection{Evaluation using Co-Purchasing Data}
\subsubsection{Method}
To prove that our results have implications for real-world business, we compared our results with the purchasing histories of customers in the Pointcard100 and Creditcard81 datasets.
We were able to create a brand-user matrix $M$ based on the co-purchasing data. 
${M}_{b_{i}u_{k}} = t$ indicates that user $u_k$ purchased brand $b_i$ $t$ times.
Then, we calculated the Pearson correlation coefficient between each pair of brands as similarity score. For example, the similarity score between brand $b_i$ and $b_j$ was calculated using vector ${M}_{b_i}$ and ${M}_{b_j}$.
Then, we compared the results based on our proposed methods with the results calculated from the co-purchasing data using the Spearman's ranking correlation.

\subsubsection{Results of Evaluation using Co-Purchasing Data}
Table~\ref{table:spearman1} shows the Spearman's ranking correlation coefficient for brand similarity based on our proposed methods and based on co-purchasing results from the point card and credit card usage. In all tables, \textit{hist} stands for the histogram-based methods; \textit{avg} stands for the average vector methods; \textit{p} stands for using the Pearson correlation; \textit{hi} stands for using histogram intersection similarity; \textit{freq} stands for ranking by frequency; \textit{score} stands for ranking by tag score.
For histogram-based methods, results changed a little when choosing different $K$ and we empirically found that using $K=500$ has better result in our proposed methods. We also compared our method with \textit{brand\_img} by \cite{gelli2018beyond}. Their method used images posted by the brand as the input and produced a brand vector that aimed to discovered images that match the brand style. We used their implementation to train the brand vector then calculate the Pearson correlation between each pair of brands as similarity score.

In the case of images, the average vector based method yields better results than the histogram-based methods. Also, proposed methods show better performance than \textit{brand\_img} \cite{gelli2018beyond}, proving that using followers' posts is better.
We could see that the tag feature works better than the image feature when compared with the co-purchasing results.
When using the histogram-based methods, histogram intersection similarity yields better results than the Pearson correlation.
With regard to the average vector based methods, ranking by tag score gives better results than ranking by frequency.
The highest Spearman's ranking correlation coefficient achieves 0.50, which is pretty high considering that this result was calculated among different groups of users (point/credit card users and brand followers are totally different persons).
The correlation coefficients of the Creditcard81 is lower than those of Pointcard100 because the users of the credit card are distributed all over the country and it is not guaranteed that all the brands have their shops within their shopping territory.

\begin{table}[t!]
	\begin{center}
		\caption{Spearman's ranking correlation coefficient between \\ proposed methods and co-purchasing results.}
		\vspace{-0.15cm}
		\label{table:spearman1}
		\begin{tabular}{|c|c|c|c|} \hline
			\multicolumn{2}{|c|}{}&\multicolumn{2}{c|}{Co-purchasing result}  \\ \cline{3-4}
			\multicolumn{2}{|c|}{}& Pointcard100 & Creditcard81 \\ \hline
			\multirow{4}{*}{Image} & brand\_img \cite{gelli2018beyond} & 0.22 & 0.19\\ \cline{2-4}
			& hist-p & 0.33 & 0.34\\ \cline{2-4}
			& hist-hi & 0.32 & 0.33\\ \cline{2-4}
			& avg-p & \textbf{0.37} & \textbf{0.37}\\ \hline
			\multirow{6}{*}{Tag}&hist-freq-p & 0.44 & 0.4\\ \cline{2-4}
			&hist-freq-hi & \textbf{0.50} & 0.43\\ \cline{2-4} 
			&hist-score-p&0.42&0.41\\ \cline{2-4}
			&hist-score-hi&0.48&0.43\\ \cline{2-4}
			&avg-freq-p& 0.44 & 0.43\\ \cline{2-4}
			&avg-score-p& 0.45 & \textbf{0.44}\\ \hline 
		\end{tabular}
		\vspace{-0.3cm}
	\end{center}
\end{table}


\subsection{Evaluation using Questionnaires}
\subsubsection{Method}
To further evaluate our results, we used questionnaires to ask users regarding their awareness of certain brands and their past purchases of the products of those brands through a crowdsourcing service.
Two separate questionnaires were asked for the Pointcard100 and Creditcard81 datasets. Nine hundred people answered the questionnaire for Pointcard100 dataset, and 890 people answered the questionnaire for Creditcard81 dataset.
For both questionnaires, we selected the top 50 brands from the dataset according to their number of followers on Instagram. The two questionnaires were similar.
We asked people three questions related to each brand.
\begin{itemize}
	\item \textit{Have you purchased this brand by yourself before?}
	\item \textit{Are you interested in this brand?}
	\item \textit{Do you know this brand?}
\end{itemize} \footnotetext{\url{https://yiwei51.github.io/brand_visualization}}
We obtained the users' co-purchasing, interest and knowledge tendencies from the responses to the questionnaires. Similar to the purchasing data in the previous section, we created three brand-user matrices, which represented the co-purchasing, interest and knowledge tendencies separately.
For example, for the first question, the response ${M}_{b_{i}u_{k}} = 1$ indicated that user $u_{k}$ had bought brand $b_{i}$ before, whereas the response ${M}_{b_{i}u_{k}} = 0$ indicated that user $u_{k}$ had never bought brand $b_{i}$ before.
Then, we calculated the Pearson correlation coefficient between each pair of brands.

\subsubsection{Results of Evaluation using Questionnaires}
Table~\ref{table:questionnaires} shows the results for the two datasets.
For Pointcard100 dataset, we can see that the Spearman's ranking correlation coefficient between the co-purchasing results obtained from the point card data and from the questionnaires is 0.52, which is high considering that co-purchasing activity is strongly affected by locality, personality, etc.
The co-purchasing results obtained from the point card data have higher correlation than the results calculated from our proposed methods. However, comparison of the customer interest obtained from the questionnaires and that from the knowledge results shows that the proposed method based on histogram has the highest Spearman's ranking correlation coefficient.
The highest correlation with the customer interest results obtained from the questionnaires is 0.53, whereas the highest correlation with the knowledge results obtained from the questionnaires is 0.50.
For Creditcard81 dataset, when compared with the customer co-purchasing, interest and knowledge results obtained from the questionnaires, the proposed methods based on the tags show stronger correlation than the co-purchasing results obtained from the credit card data.

From these results, it can be said that we have grasped the tendency of customer preferences. Our proposed methods can predict customers' interests more accurately than the predictions made from the customer's purchasing history. 
We considered the reason for this result might be related to age or finances.
For example, women in their early twenties might be unable to purchase some high-end fashion brands because of age-related and insufficient finances, but they might be interested in these brands to prepare for their late twenties 
or thirties. We believe that our proposed methods could reflect a customer's future purchasing plan.

\begin{table*}[t!]
	\begin{center}
		\caption{Spearman's ranking correlation coefficient between the proposed methods and questionnaires.}
		\vspace{-0.3cm}
		\label{table:questionnaires}
		\begin{tabular}{|c|c|c|c|c||c|c|c|} \hline
			\multicolumn{2}{|c|}{}&\multicolumn{3}{c||}{Pointcard100 Dataset}&\multicolumn{3}{c|}{Creditcard81 Dataset} \\ \cline{3-8}
			\multicolumn{2}{|c|}{}& Purchase & Interest & Knowledge & Purchase & Interest & Knowledge\\ \hline
			\multirow{4}{*}{Image} & brand\_img \cite{gelli2018beyond} &0.20&0.24&0.18&0.23&0.25&0.14\\ \cline{2-8}
			& hist-p &0.26&0.34&0.28&0.20&0.31&0.15\\ \cline{2-8}
			& hist-hi &0.26&0.38&0.33&0.23&0.32&0.17\\ \cline{2-8}
			& avg-p &0.27&0.39&0.34&0.25&0.35&0.18\\ \hline 
			\multirow{6}{*}{Tag}&hist-freq-p &0.36&\textbf{0.53}&0.49&0.31&0.40&0.24\\ \cline{2-8}
			&hist-freq-hi &0.36&\textbf{0.53}&0.49&\textbf{0.34}&0.42&\textbf{0.26}\\ \cline{2-8} 
			&hist-score-p&0.33&\textbf{0.53}&\textbf{0.50}&0.31&0.40&0.22\\ \cline{2-8}
			&hist-score-hi&0.36&0.52&\textbf{0.50}&\textbf{0.34}&\textbf{0.43}&\textbf{0.26}\\ \cline{2-8}
			&avg-freq-p&0.32&0.51&0.49&0.30&0.39&0.22\\ \cline{2-8}
			&avg-score-p&0.34&0.50&0.48&0.31&0.41&0.22\\ \hline \hline
			\multicolumn{2}{|c|}{Co-purchasing result obtained by point/credit card data}&\textbf{0.52}&0.23&0.21&0.28&0.33&0.17\\ \hline
		\end{tabular}
	\vspace{-0.2cm}
	\end{center}
\end{table*}

\section{Applications}

We have introduced that different brands' followers might have different interest tendencies as shown in Figure~\ref{fig:brand_follower}.
Based on followers' interest tendencies, we could recommend suitable content to the brand for popularity enhancement.
We cooperated with a fast food company from the end of 2018.
We provided results of analysis to the brand including tags ranking by tag score, similar and dissimilar brands to it. And the fast food company changed their posts based on our recommendations.
We recorded the number of favorites of every post from that fast food company's official account.
The average number of favorites is 477 from 2015 to 2018, and increased to 3,788 in 2019. The number of favorites suddenly increased after the consulting with us proves that our proposed methods could be a novel marketing tool for popularity boosting.


\section{Conclusions}

In this paper, we proposed a new scale for measuring the similarity between brands using image features and tag features of social media posts made by their followers. 
Then, we evaluated our results by comparing with the results of the co-purchasing history of real-world customers and the results from the questionnaires. We found that our proposed methods have moderate correlations with the customers' co-purchasing histories. In addition, we found that results based on our proposed methods show stronger correlation with users' interest tendencies obtained from questionnaires than results obtained from the customers' co-purchasing histories.
We believe that this multimedia-based follower analysis in social networks can become a novel marketing tool. 

\begin{acks}
This work was partially supported by the Grants-in-Aid for Scientific Research Number JP19J22939, JP19K20289, and JP18H03339. A part of research is a result of collaboration between CyberBuzz, Inc. and The University of Tokyo.
\end{acks}

%
\bibliographystyle{ACM-Reference-Format}
\bibliography{sample-base}

\end{document}